\documentstyle[aps,epsf,floats]{revtex}
\begin{document}

\begin{flushright}
JLAB-THY-98-41 \\
October 23,1998\\
\end{flushright}

\begin{center}
{\Large \bf Symmetries and 
Structure of Skewed and  Double  Distributions   }
\end{center}
\begin{center}
{A.V. RADYUSHKIN\footnotemark
}  \\
{\em Physics Department, Old Dominion University,}
\\{\em Norfolk, VA 23529, USA}
 \\ {\em and} \\
{\em Jefferson Lab,} \\
 {\em Newport News,VA 23606, USA}
\end{center}
\vspace{2cm}

\footnotetext{Also Laboratory of Theoretical  Physics, 
JINR, Dubna, Russian Federation}

\begin{abstract}

Extending the concept of 
 parton densities onto   nonforward matrix
elements $\langle p' \,  |     \,   
{\cal O}(0,z)  \,  |     \, p \rangle $ 
of quark and gluon light-cone  operators,
one can  use
two types of nonperturbative functions:
double distributions (DDs) $f(x,\alpha;t), F(x,y;t)$ 
and  skewed (off\&nonforward) parton 
 distributions (SPDs) $H(\tilde x,\xi;t),
  {\cal F}_{\zeta}(X,t)$.   
  We treat DDs as  primary objects 
  producing SPDs after  integration. 
We emphasize the role of DDs 
in understanding interplay between $X$ ($\tilde x$)
and $\zeta$ ($\xi$) dependences of SPDs.
In particular, the use of DDs 
is crucial to secure the polynomiality 
condition:  $N$th moments of SPDs are $N$th
degree polynomials in the relevant 
skewedness parameter $\zeta$ or $\xi$.
We propose   simple ans\"atze  for  DDs 
having  correct  
spectral and symmetry properties and 
derive  model expressions for 
SPDs satisfying  all  known constraints. 
Finally, we argue that for small skewedness,
one can obtain SPDs from the usual 
parton densities by averaging the latter 
with an appropriate weight over the region
$[X-\zeta,X]$ (or $[\tilde x - \xi,\tilde x + \xi]$).

\vspace{5mm}

PACS number(s): 12.38.Bx, 13.60.Fz, 13.60.Le

\end{abstract}

%\newpage

\section{Introduction}

Nonforward matrix
elements $\langle p-r \,  |     \,   
{\cal O}(0,z)  \,  |     \, p \rangle \,  |     \, _{z^2=0}$ 
of quark and gluon light-cone  operators
which appear in applications of perturbative QCD to
deeply virtual Compton scattering (DVCS) 
and hard exclusive electroproduction 
processes \cite{ji,compton,npd,cfs,drm} 
  can be parametrized by 
two basic  types of nonperturbative functions. 
The   double distributions (DDs) $F(x,y;t)$ \cite{compton,npd} 
specify the Sudakov
light-cone ``plus'' fractions $xp^+$ and $yr^+$ 
of the initial hadron momentum $p$ and the momentum transfer $r$ 
carried by the initial parton.   
Treating the proportionality  coefficient 
$\zeta$ as an 
independent parameter  one can   introduce 
an alternative description in terms
of the   nonforward parton distributions (NFPDs) 
${\cal F}_{\zeta}(X;t)$ 
with $X=x+y \zeta$ being the total 
fraction of the initial hadron momentum 
taken  by the initial  parton.
The shape of  NFPDs  explicitly
depends on the parameter $\zeta$ characterizing the {\it skewedness}
of the relevant nonforward matrix element.
This parametrization of  nonforward matrix 
elements  by ${\cal F}_{\zeta}(X;t)$
is similar to that proposed  by 
X. Ji \cite{ji} who introduced 
 off-forward parton distributions (OFPDs) $H(\tilde x,\xi;t)$
in which the parton momenta and  the skewedness
parameter $\xi\equiv r^+ / P^+$ 
are measured in units of the average 
hadron momentum $P=(p+p')/2$. 
There are one-to-one  relations between  OFPDs and NFPDs \cite{npd},
so   
it is convenient to treat 
  them 
as particular forms  of {\it skewed } parton 
distributions (SPDs).

In our approach, DDs are  primary
objects producing SPDs after an appropriate integration.
Our main goal in this letter is to show that  
using the  formalism of  DDs (in particular,  their 
support and symmetry properties) 
one can easily establish important features of SPDs
such as nonanalyticity at border points $X=\zeta,0$ and 
$\tilde x = \pm \xi$, polynomiality of their $X^N$ and 
$\tilde x^N$ moments in skewedness parameters $\zeta$ and $\xi$, etc.
We also discuss simple models for DDs which 
result in realistic models for SPDs.

\section{Double distributions and their symmetries}

\begin{figure}[htb]
\mbox{
   \epsfxsize=13.5cm
 \epsfysize=7.5cm
 \hspace{1.5cm}  
  \epsffile{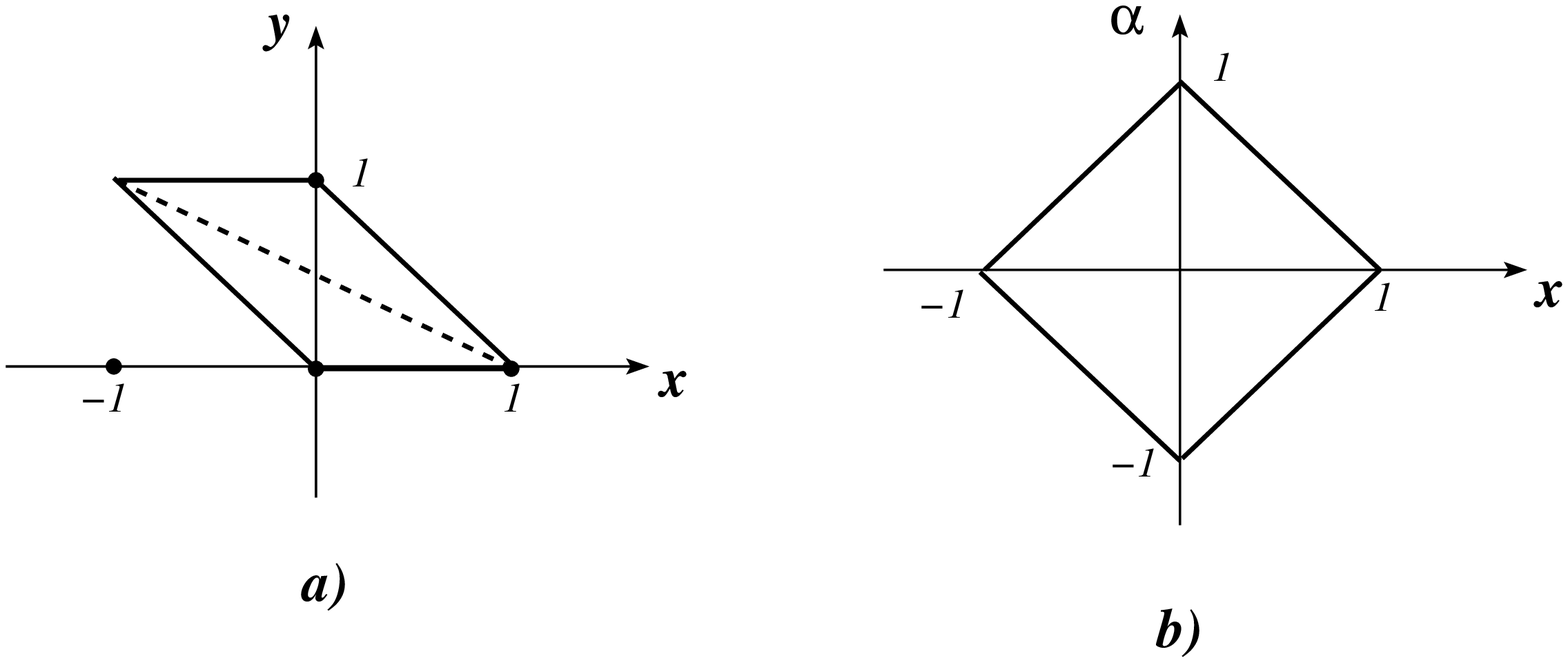}  }
  \vspace{-1.5cm} 
{\caption{\label{fg:support} $a)$ Support region 
and symmetry line  $y = \bar x/2$ for 
$y$-DDs 
$\tilde F(x,y;t)$; $b)$ support region 
for $\alpha$-DDs $\tilde f (x, \alpha)$. 
   }}
\end{figure}

In the pQCD factorization treatment of hard
electroproduction processes, the 
nonperturbative information is accumulated in the
nonforward matrix elements
$\langle p-r \,  |     \,  {\cal O} (0,z)  \,  |     \,  p \rangle $
of   light cone operators $ {\cal O} (0,z) $. 
For  $z^2=0$ the matrix elements 
depend on the relative coordinate $z$
through two Lorentz invariant variables $(pz)$ and $(rz)$.
In the forward case, when $r=0$, 
one obtains the usual quark helicity-averaged densities   
by  Fourier transforming   the relevant  matrix element 
with respect to $(pz)$
\begin{equation} 
\langle p,s'\, \,  |     \,  \, \bar \psi_a(0) \hat z 
E(0,z;A)  \psi_a(z) \, \,  |     \,  \, p,s \rangle \,  |     \, _{z^2=0} 
 =  \bar u(p,s')  \hat z u(p,s)  
   \int_0^1  \,  
 \left ( e^{-ix(pz)}f_a(x) 
  -   e^{ix(pz)}f_{\bar a}(x)
\right ) \, dx \, , 
\label{33} \end{equation} 
where $E(0,z;A)$ is the gauge link,  
 $\bar u(p',s'), u(p,s)$ are the Dirac
spinors and we use the  notation
$\gamma_{\alpha} z^{\alpha} \equiv \hat z$.
In the nonforward case,  we can  
use the  double Fourier  representation 
with respect to both $(pz)$ and $(rz)$: 
\begin{eqnarray} 
&& \langle p',s'\, \,  |     \,  \, \bar \psi_a(0) \hat z 
E(0,z;A)  \psi_a(z) \, \,  |     \,  \, p,s \rangle \,  | 
    \, _{z^2=0} 
\label{31}  \\ &&  =  \bar u(p',s')  \hat z u(p,s)  
\int_0^1  dy   \int_{-1}^1  \,  
  e^{-ix(pz)-iy(r z)} \, \tilde F_a(x,y;t) \, 
 \theta( 0 \leq x+y \leq 1) \, dx  \, + \, ``\tilde K"-{\rm term}  \, , 
\nonumber 
 \end{eqnarray} 
where  the
``$\tilde K$''-term stands for  the hadron helicity-flip   part \cite{ji,npd}. 
 For any Feynman 
diagram, the spectral constraints 
$-1 \leq x \leq 1$, $0 \leq y \leq 1$, $0 \leq x+y
\leq 1$ were  proved in the $\alpha$-representation
\cite{npd} using the approach of Ref.
\cite{spectral}. The support area 
for  the {\it double distribution} $ \tilde F_a(x,y;t)$ 
is shown on Fig.\ref{fg:support}a. 

\vspace{-1cm}

Comparing Eq. (\ref{33}) 
with the $r =0$ limit of the DD definition (\ref{31})
gives  the   
``reduction formulas'' 
relating   the double distribution $\tilde F_{a}(x,y;t=0)$  
to the quark and antiquark parton  densities
\begin{equation} 
\int_0^{1-x} \, \tilde F_{a}(x,y;t=0)|_{x>0} \, dy= 
 f_{a}(x) \hspace{1cm} ; \hspace{1cm} 
\int_{-x}^1 \, \tilde F_{ a}(x,y;t=0)|_{x< 0}\, dy= 
-  f_{\bar a}(-x) \,  \label{eq:redfsym} \, .
 \end{equation}
Hence, the  positive-$x$
and negative-$x$  components of the double 
distribution $\tilde F_{a}(x,y;t) $
can be treated as nonforward generalizations of 
quark and antiquark densities, respectively.
If we define the ``untilded'' DDs by 
\begin{equation} 
F_{a}(x,y;t) = \tilde F_{a}(x,y;t)|_{x>0}  
\hspace{1cm} ; \hspace{1cm} 
F_{\bar a}(x,y;t) = - \tilde F_{a}(-x,1-y;t)|_{x<0}   \, ,
\label{abara}   \end{equation}
then $x$ is always positive and 
 the reduction formulas  have  the same form
\begin{equation} 
\int_0^{1-x} \, F_{a,\bar a}(x,y;t=0)|_{x \neq 0} \, dy= 
 f_{a,\bar a}(x) 
 \label{34} \end{equation}
in both cases. The new antiquark distributions  
also ``live'' on the triangle
$0 \leq x,y \leq 1, \, 0 \leq x +y \leq 1$.
Taking $z$ in the lightcone ``minus''  direction,
we arrive at  the   parton interpretation 
of  functions $ F_{a, \bar a} (x,y;t )$ as probability amplitudes for an 
outgoing  parton to carry the fractions $xp^+$
and $yr^+$ of the external 
momenta $r$ and $p$. 
The  double distributions 
$F(x,y;t)$  are universal functions 
describing the flux of $p^+$ and $r^+$ 
independently of the ratio $r^+/p^+$. 
Note, that  extraction  of two separate
components $F_a(x,y;t)$ and  $F_{\bar a}(x,y;t)$ 
from the quark DD  $\tilde F_a(x,y;t)$ as its 
positive-$x$ and negative-$x$ parts  is unambiguous.

\begin{figure}[htb]
\mbox{
   \epsfxsize=12cm
 \epsfysize=5cm
 \hspace{0.5cm}  \epsffile{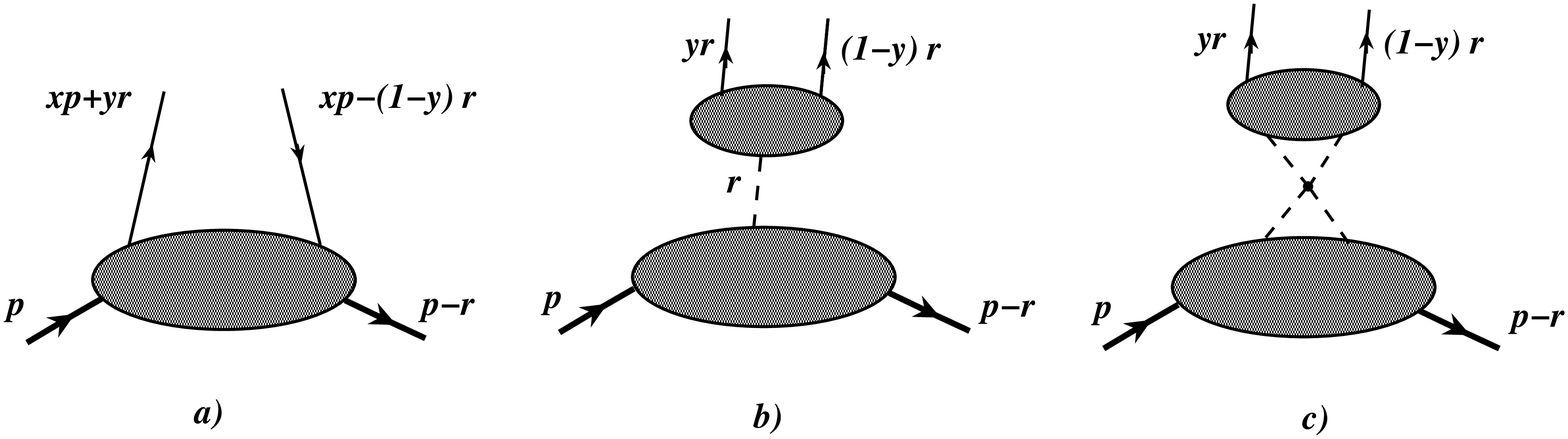} 
 \hspace{-1cm}  \epsfxsize=5cm
 \epsfysize=5cm \epsffile{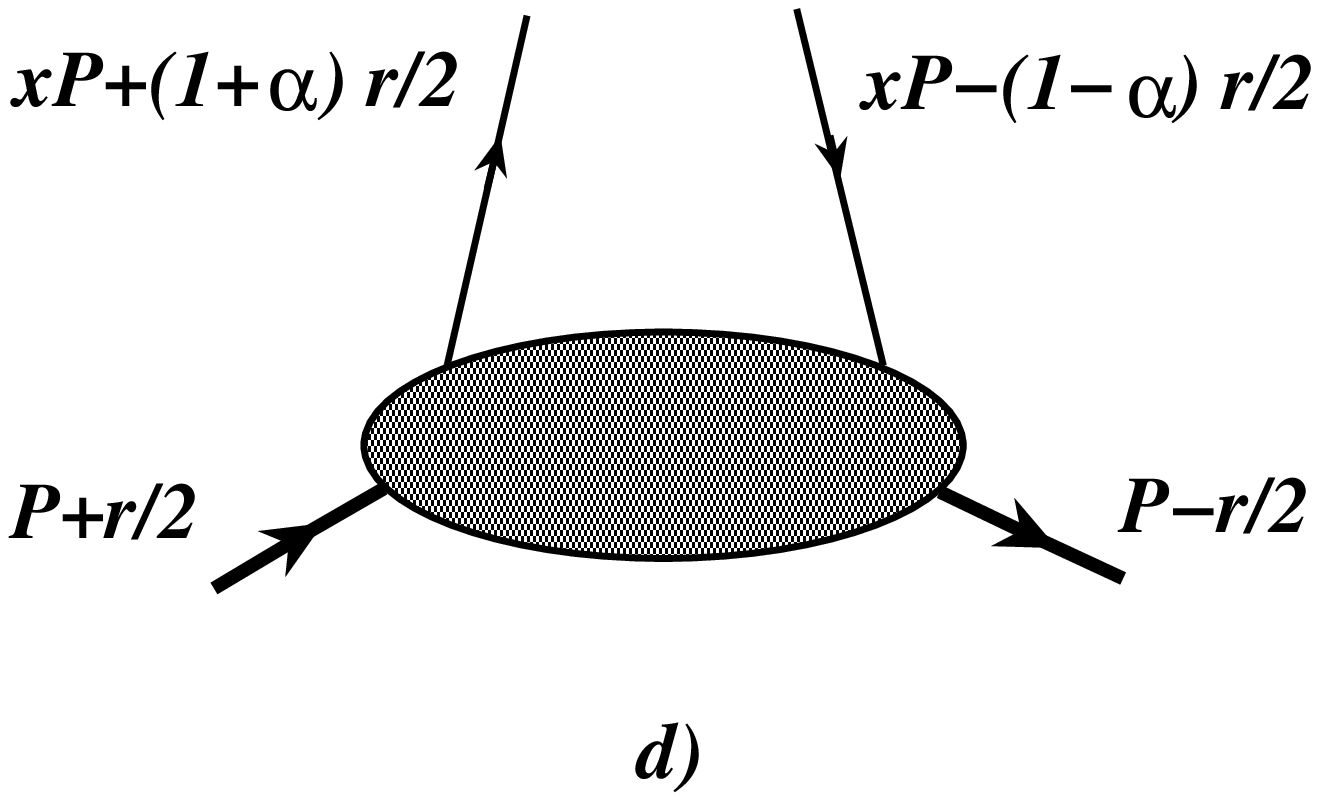} 
   }
  \vspace{-1.5cm}
{\caption{\label{fg:exchange} $a)$ Parton picture
in terms of $y$-DDs; $b,c)$   $F_M$-type contributions;
$d)$ parton picture
in terms of $\alpha$-DDs.
   }}
\end{figure}

In principle, we cannot exclude 
the  third possibility that  the functions 
$\tilde F (x,y;t)$    have  singular terms at $x=0$ 
proportional 
to $\delta (x)$ or its derivative(s).
Such terms  
have no projection  onto the usual parton densities.
We will denote them by $F_M (x,y;t)$ $-$  they may 
be interpreted as coming from the $t$-channel 
meson-exchange type contributions (see Fig.\ref{fg:exchange}b).  
In this case, the partons  just share 
the plus component of the momentum transfer $r$:
information about 
the magnitude of the  initial hadron momentum
is lost if the exchanged particle can be described
by a pole propagator $\sim 1/(t-m_M^2)$.
Hence, the meson-exchange  contributions to a double distribution
may look like
\begin{equation}
\tilde F_M^+(x,y;t) \sim  \delta (x) \,
 \frac{\varphi_{M}^+ (y)}{m_M^2 -t } \ \  \ 
{\rm or }  \ \  \  \tilde  F_M^-(x,y;t) 
\sim  \delta ' (x) \, \frac{\varphi^-_{M
}(y)}{m_M^2 -t } \ \  , \  \  {\rm etc.} \, , 
\end{equation} 
where $\varphi_M^{\pm}  (y)$ are  the functions related to the 
distribution amplitudes of the relevant mesons $M^{\pm}$.
The  two examples above correspond to $x$-even and $x$-odd
parts of the double distribution $\tilde F(x,y ;t)$. 
The  singular terms can also be produced by 
diagrams containing  
a quartic pion vertex (Fig.\ref{fg:exchange}c) \cite{christian}.

To make the description more symmetric 
with respect to the  initial and final hadron momenta,  
we can  treat nonforward matrix elements as 
functions of $(Pz)$ and $(rz)$, where $P=(p+p')/2$
is the average hadron momentum. 
The relevant  double distributions 
$\tilde f_a(x,\alpha\,;t)$ [which we 
will call $\alpha$-DDs to distinguish them
from $y$-DDs $F(x,y;t)$] are defined by  
\begin{eqnarray}
\langle p' |  \bar \psi_a (-z/2) \hat z
\psi_a (z/2) |p \rangle \, = 
\bar u(p') \hat z u(p) \, \int_{-1}^1 \, dx \int_{-1+|x|}^{1-|x|} 
 e^{-ix(Pz)-i\alpha (rz)/2}  \, \tilde f_a(x,\alpha;t) 
 \,
d\alpha  \,   
 + ``\tilde k"{\rm -terms} \, .
 \label{17} \end{eqnarray}
The support area for $\tilde f_a(x,\alpha;t)$ is shown 
in Fig.\ref{fg:support}b.
Again, the usual forward densities 
$f_a(x)$ and $f_{\bar a} (x)$ 
are given by integrating 
$\tilde f_a(x,\alpha\,;\,t=0)$  over vertical lines
$x = {\rm const} $ for $x>0$ and $x<0$, respectively.
Hence, we can  split $\tilde f_a(x,\alpha\, ;\,t)$ 
into  three components 
\begin{equation} 
\tilde f_a(x,\alpha\,;\,t) = f_a(x,\alpha\,; \,t)\, \theta (x>0) - 
f_{\bar a} (-x, - \alpha\,; \,t)\, 
\theta (x<0) + f_M (x, \alpha\,; \,t) \, ,
\label{fff} \end{equation}
where $f_M (x, \alpha\,; \,t)$ is a singular  term
with support at $x=0$ only\footnote{As argued by 
M. Polyakov and C. Weiss \cite{christian}, 
in the case of   pion distributions it makes sense to
  write the $(Pz)$-independent
terms as a separate integral  over a single variable 
$y$ rather than to
include them into  a singular part of DDs.}.  Due  to hermiticity
and time-reversal invariance properties of
nonforward matrix elements,  the $\alpha$-DDs
are even functions of $\alpha$: 
$$
\tilde f_a(x,\alpha;t) = \tilde f_a(x, - \alpha;t) \, . 
$$
For our original $y$-DDs $F_{a, \bar a} (x,y;t)$, this 
corresponds to symmetry with respect to  the interchange 
$y \leftrightarrow 1-x-y$ established in Ref. \cite{lech}. In
particular, the  functions  $\varphi_M^{\pm}(y)$ for singular contributions
$F_M^{\pm}(x,y;t)$ are  symmetric  
$\varphi_M^{\pm}(y) = \varphi_M^{\pm}(1-y) $
both for  $x$-even and $x$-odd parts.
The  $a$-quark contribution 
$$
{\cal O}_{ a}^S(-z/2,z/2) =  
 \frac{i}{2}
[ \bar \psi_a(-z/2) 
\hat z E(-z/2,z/2;A)  \psi_a(z/2)
- \{ z \to -z \}  ]   
  $$
into the 
 flavor-singlet 
operator
can be  parametrized either by  $y$-DDs $\tilde F_a^S (x,y;t)$ or  
by $\alpha$-DDs $\tilde f_a^S(x, \alpha \, ; \, t)$  
\begin{eqnarray} 
&& \langle \,  p',s' \,  |     \,  \, {\cal O}_{ a}^S(-z/2,z/2)\, 
\,  |     \,  \, p,s \rangle \,  |     \, _{z^2=0} \nonumber \\  &&  
= 
 \bar u(p',s')  \hat z  u(p,s)   
\int_0^1  dx \int_0^{1-x}   \, 
  \frac1{2} \left( e^{-ix(pz)-i(y-1/2)(r z)} 
- e^{ix(pz)+i (y-1/2) (r z)}\right ) 
\,  F_a^S(x,y;t)  
  \, dy  + ``K_a^S"{\rm -term} \nonumber 
\\  &&  
= \bar u(p',s')  \hat z 
 u(p,s) \,  \int_{-1}^1  dx  \int_{-1+|x|}^{1-|x|}  \, 
    e^{-ix(Pz)-i\alpha (r z)/2} 
\,  \tilde f_a^S(x,\alpha \, ; \, t)  
 \, d\alpha  + ``\tilde k_a^S"{\rm -term}
.
\label{311Q} \end{eqnarray} 
In the second line here we have used the fact that 
  positive-$x$ and negative-$x$  parts 
  in this case 
are described by the  same untilded function  
$$
  F_a^S(x,y;t)|_{x \neq 0}   =  
 F_a(x,y;t) + F_{\bar a} (x,y;t)  .
$$
The  $\alpha$-DDs $\tilde f_a^S(x, \alpha \, ; \, t)$
are even functions of   $\alpha$ and, according to
Eq. (\ref{311Q}),    odd functions of $x$:
\begin{equation}
  \tilde f^{S}_a (x,\alpha ;t)   =   
  \{ f_a(|x|,|\alpha | ;t) + f_{\bar a} (|x|,|\alpha| ;t) \} \, 
  {\rm sign} (x) +f_M^S(x,\alpha ;t) \ .
\label{singlet} \end{equation} 
Finally, the valence 
 quark functions $\tilde f_a^{V}(x, \alpha \, ; \, t)$
related to the operators 
$$  {\cal O}^V_a (-z/2,z/2) = \frac12 [\bar \psi_a(-z/2) 
\hat z  E(-z/2,z/2;A)   \psi_a(z/2)
+  \{ z \to -z \} ]$$
are even functions of both $\alpha$  and $x$:
\begin{equation}
  \tilde f^{V}_a (x,\alpha ;t)   =   
  f_a(|x|,|\alpha| ;t) - f_{\bar a} (|x|,|\alpha|;t) + f_M^V(x,\alpha ;t) \ .
\label{valence} \end{equation}

\section{Parton interpretation 
and models for double distributions}

\begin{figure}[htb]
\mbox{
   \epsfxsize=14cm
 \epsfysize=8cm
 \hspace{1cm}  
  \epsffile{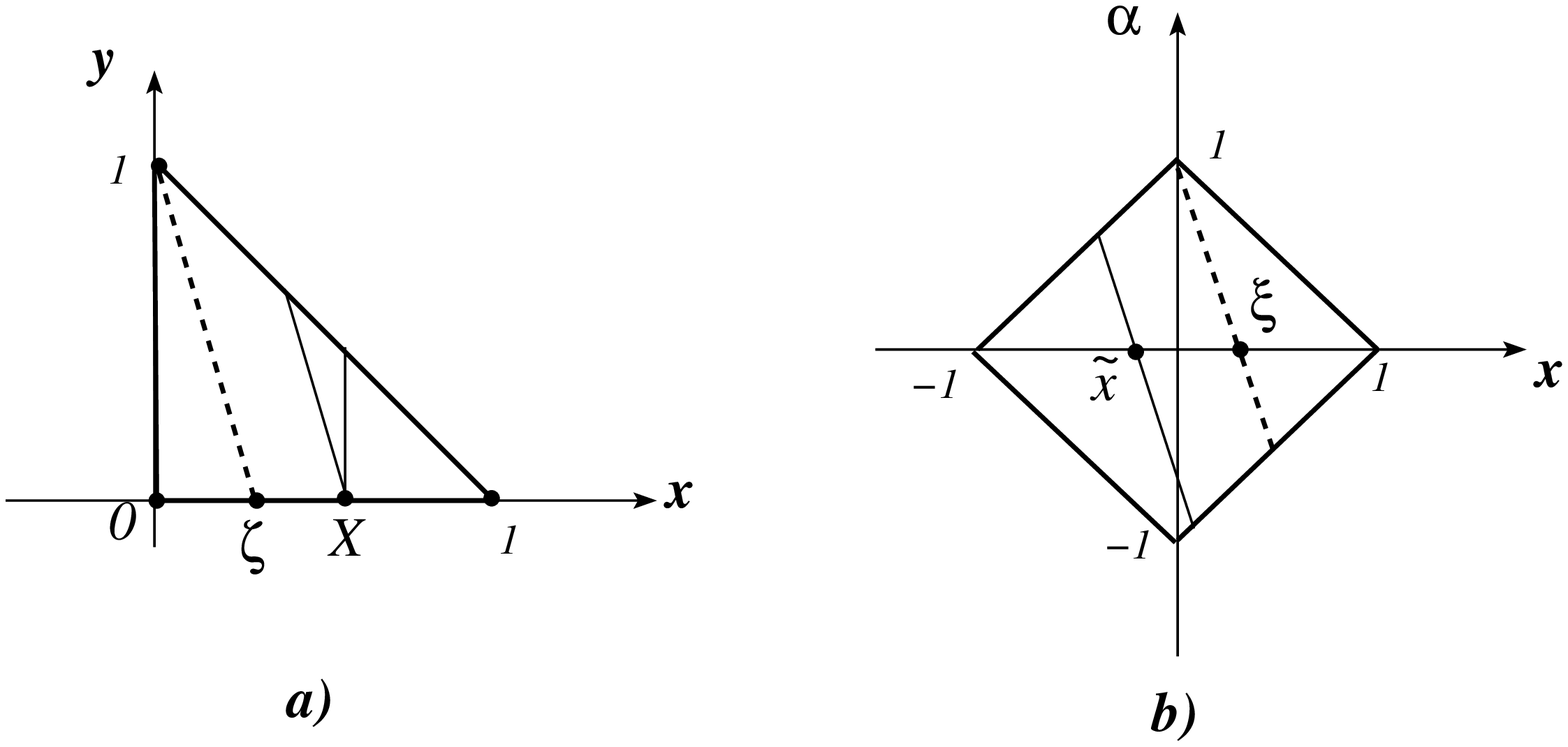}  }
  \vspace{-1cm}
{\caption{\label{fg:spds}    Integration lines 
for integrals relating SPDs and DDs. 
   }}
\end{figure}

The structure of the integral  (\ref{34}) relating 
double distributions with the usual ones 
 has a simple graphic    illustration
(see Fig.\ref{fg:spds}): 
 integrating  DDs 
over a line orthogonal to the $x$  axis, we get $f(X)$.

The  reduction formulas and   interpretation of
the  $x$-variable  as the  fraction of 
the  $p$ (or $P$) momentum 
suggest that the   profile of $F(x,y)$ (or $f(x,\alpha)$)  
in  $x$-direction is basically determined by the shape 
of $f(x)$. 
On the other hand, the profile in  $y$ (or $\alpha$) direction  
characterizes the spread of the parton momentum induced by
the momentum transfer $r$. 
In particular, since 
the $\alpha$-DDs $f(x,\alpha)$ 
are even functions of $\alpha$,
it make sense to write 
\begin{equation}
f(x,\alpha) =  h(x,\alpha) \,  f(x)  \, ,  \label{65n}
 \end{equation}
 where $h(x,\alpha)$ is an even function of $\alpha$ 
 normalized by 
\begin{equation}
 \int_{-1+x}^{1-x} h(x,\alpha) \, d\alpha \, =1.
 \end{equation}
We may expect that 
the $\alpha$-profile of $h(x,\alpha)$  
is similar to that of a symmetric distribution amplitude (DA) 
$\varphi (\alpha)$.  Since $|\alpha| \leq \bar x $, to get a 
more complete  analogy
with DAs, 
it makes sense to rescale $\alpha$ as $\alpha = \bar x \beta$
introducing  the  variable $\beta$ with $x$-independent limits:
$-1 \leq \beta \leq 1$. 
 The simplest model is to assume 
that the profile in the $\beta$-direction is  
 a  universal function  $g(\beta)$ for all $x$. 
Possible simple choices for  $g(\beta)$ may be  $\delta(\beta)$
(no spread in $\beta$-direction),  $\frac34(1-\beta^2)$
(characteristic shape for asymptotic limit 
of nonsinglet quark distribution amplitudes), 
 $\frac{15}{16}(1-\beta^2)^2$
(asymptotic shape of gluon distribution amplitudes), etc.
In the variables $x,\alpha $, this gives   
\begin{equation}
h^{(0)} (x,\alpha) =  \delta(\alpha) \,   \ , \
h^{(1)}(x,\alpha) = \frac{3}{4} 
\frac{ (\bar x^2 - \alpha^2)}{(1-x)^3} \,   \ , \
h^{(2)}(x,\alpha) = \frac{15}{16} 
\frac{(\bar x^2 - \alpha^2)^2}{(1-x)^5} \,   \  . \label{mod123} 
 \end{equation}
 It is straightforward to  generalize
 these models onto the 
 ``tilded'' DDs  $\tilde f (x,\alpha)$ with
 $x$ ranging between $-1$ and $1$:
 $\tilde f (x,\alpha)$ should be  even in $x$
 for the gluon and nonsinglet quark distributions
 and odd in $x$ for   the singlet quark case.
 Furthermore, one can construct ans\"atze 
for functions $f(x,\alpha;t)$ involving  nonzero $t$ values, e.g., 
the model 
\begin{equation}
f_i(x,\alpha;t ) =  h(x,\alpha) \,  
f_i (x) \exp \left \{ \frac{(\bar x^2 - \alpha^2)t}{4 x \bar x
\lambda^2} \right  \}  \label{65}  \end{equation} 
 with  $h(x,\alpha) = \delta (\alpha)$ 
and experimental valence densities $f_{u,d}^{V}(x)$
was used in ref.\cite{realco} to describe the 
$F_1 (t)$ form factor and wide-angle 
Compton scattering.

\section{Relations between double and skewed  distributions}

An important parameter for nonforward matrix 
elements is the  coefficient of proportionality 
$\zeta = r^+/p^+$  (or $\xi = r^+/P^+$)  between 
the plus components of the momentum transfer 
and initial (or  average) hadron  momentum. 
It  specifies 
the {\it skewedness}  of the matrix elements.
The characteristic   feature  implied by
 representations for double distributions 
 [see, e.g., Eq.(\ref{31})]
 is the absence 
of the $\zeta$-dependence in 
the DDs   $F(x,y)$ and $\xi$-dependence in $f(x,\alpha)$.
An alternative way to parametrize 
nonforward matrix elements of light-cone operators
is to use   
 $\zeta$  (or $\xi$) and the {\it total } 
 momentum fractions  $X \equiv x+y \zeta$ 
(or $\tilde x \equiv  x +\xi \alpha$) 
  as  independent 
variables. 
If we require that 
the light-cone plus components  
of both the momentum transfer
$r$ and the final hadron momentum  $p - r$ are 
positive, then  $0 \leq \zeta \leq 1$ and $0 \leq \xi \leq 1$.
Using the spectral property   $0 \leq x+y \leq 1$
of  double distributions,  we obtain that 
the NFPD variable $X$ satisfies the  ``parton''   
constraint $0\leq X \leq 1$.
Integrating each particular  double   distribution 
over $y$ gives  the nonforward  parton distributions \cite{npd} 
\begin{eqnarray} && 
{\cal F}_{\zeta}^{i} (X) = \int_0^1 dx \int_0^{1-x} 
\, \delta (x+\zeta y -X) \, F_i(x,y) \, dy \nonumber \\ &&
=
\theta(X \geq \zeta) 
 \int_0^{ \bar X / \bar \zeta } F_{i}(X-y \zeta,y) \, dy + 
 \theta(X \leq \zeta)  \int_0^{ X/\zeta} F_{i}
 (X-y \zeta,y) \, dy \,  , 
\label{71}  \end{eqnarray}
where $\bar \zeta \equiv 1- \zeta$.
The two components of NFPDs correspond to positive
($X> \zeta$) and negative ($X< \zeta$)
values of the fraction $X' \equiv X - \zeta$ 
associated with the ``returning'' parton.
As explained in refs. \cite{compton,npd},
the second component  can be interpreted as the
probability amplitude for the initial hadron with momentum 
$p$ to split into the final hadron with momentum $(1-\zeta)p$
and  a two-parton state with total momentum $r=\zeta p$
shared by the partons
in fractions $Yr$ and $(1-Y)r$, where $Y=X/\zeta$.

The relation between NFPDs
and  DDs  can be     illustrated
on the ``DD-life triangle'' 
defined by $0 \leq x,y,x+y \leq 1$ (see Fig.\ref{fg:spds}a).
Specifically, to get ${\cal F}_{\zeta} (X)$,  one should 
integrate $F(x,y)$ over $y$ along a straight
line  $x=X- \zeta y$. Fixing some value of  $\zeta$,
one deals with  a set of parallel lines  intersecting  the $x$-axis
at $x=X$. The upper limit of the $y$-integration
is determined by intersection of this line
either with the line $x+y=1$ (this happens if 
$X > \zeta$)  or with the $y$-axis (if $X < \zeta$).
The line corresponding to $X=\zeta$
separates the triangle into two parts
generating the  two components of the
nonforward parton distribution.

In a similar way, we can write the relation 
between OFPDs 
$\tilde H(\tilde x,\xi;t)$ \cite{ji} 
 and the 
$\alpha$-DDs $\tilde f(x,\alpha;t)$ 
\begin{equation} 
 \tilde H (\tilde x,\xi; t)=  \int_{-1}^1 dx \int_{-1+|x|}^{1-|x|}  
\, \delta (x+ \xi \alpha - \tilde x) \, 
\tilde f (x,\alpha;t) \, d\alpha  \, . \label{offfor} 
 \end{equation}
We use here the tilded notation  $\tilde H (\tilde x,\xi; t)$ 
to emphasize that OFPDs as defined by 
X. Ji \cite{ji} correspond to parametrization
of the nonforward matrix element by a Fourier integral 
with a single  common exponential. 
Note that Eq. (\ref{offfor}) allows to construct $\tilde H (\tilde x,\xi; t)$
both for positive and negative values of $\xi$.
Since DDs $\tilde f (x,\alpha;t)$ are even functions 
of $\alpha$, the OFPDs  
$\tilde H (\tilde x,\xi; t)$ are even functions 
of $\xi$: 
$$
\tilde H(\tilde x,\xi;t) = \tilde H(\tilde x,- \xi;t )  \, .
$$
This result was originally obtained by X. Ji \cite{jirev} 
who  directly used hermiticity and time reversal
invariance  properties in his definition of OFPDs.

The  delta-function in Eq.(\ref{offfor}) specifies 
the line of  integration in the
$\{ x, \alpha \}$ plane (see Fig.\ref{fg:spds}).  
For definiteness,  we will assume below
that 
$\xi$ is positive.    The integration line $x = \tilde x - \xi \alpha$
consists of  two parts corresponding to 
positive and negative values
of $x$.  In the case of quarks with flavor
$a$,  substituting $\tilde f_a (x,\alpha)$ 
by $f_a (x,\alpha)$ or $f_{\bar a} (x,\alpha)$, respectively  
 (see Eq.(\ref{fff})), we get  
 OFPD $\tilde H_a(\tilde x,\xi;t)$
as the sum of  three components
\begin{equation}
\tilde H_a(\tilde x,\xi;t) = 
 H_a(\tilde x,\xi;t)\, \theta (-\xi \leq \tilde x \leq 1) - 
H_{\bar a} (- \tilde x,\xi;t)\, \theta (-1 \leq \tilde x \leq \xi) + 
H_M(\tilde x,\xi;t)\, \theta (-\xi
 \leq \tilde x \leq \xi) \, ,  \label{HHH}
\end{equation}
where $H_M(\tilde x,\xi;t)$ comes from integration of 
the singular term $f_{M} (\tilde x - \xi \alpha , \alpha)$ over 
$\tilde x/\xi - \epsilon <\alpha < \tilde x/\xi + \epsilon$ 
and 
\begin{equation}
H_{a, \bar a} (\tilde x,\xi;t) = 
\theta(\xi \leq \tilde x \leq 1) 
 \int_{-\frac{1- \tilde x}{1+\xi} }^{\frac{1- \tilde x}{1-\xi} }
f_{a, \bar a} (\tilde x - \xi \alpha,  \alpha ) \, d \alpha 
+ \theta(-\xi \leq \tilde x \leq \xi) 
 \int_{-\frac{1- \tilde x}{1+\xi} }^{\tilde x / \xi -\epsilon} 
f_{a, \bar a} (\tilde x - \xi \alpha,  \alpha ) \, d \alpha \, .
\label{710}  
 \end{equation}

The  OFPD $\tilde H_{a} (\tilde x,\xi;t)$ is  in a one-to-one 
correspondence with  the ``tilded'' NFPD 
$\tilde {\cal F}^a_{\zeta}  (X)$ introduced in our paper 
\cite{npd}. It 
parametrizes the nonforward  matrix element 
of the   quark operator $\bar
\psi_a \ldots \psi_a$  through  a Fourier 
integral with a single common exponential.
The  support of $\tilde {\cal F}^a_{\zeta}  (X)$ is 
$-1 +\zeta \leq X \leq 1$ and     by 
\begin{equation}
\tilde {\cal F}^a_{\zeta}  (X) = 
{\cal F}^a_{\zeta}  (X)\,  \theta (0 \leq X \leq 1) -
{\cal F}^{\bar a}_{\zeta}  (\zeta - X) 
\,\theta (-1 +\zeta  \leq X \leq \zeta)
+ {\cal F}^{M}_{\zeta}  (X) \,
 \theta (0\leq X \leq \zeta) \,  \label{three}
\end{equation}
it is  related to the  untilded     
components  given 
by Eq. (\ref{71}).  
In the middle region $0 \leq X \leq \zeta$, 
the components ${\cal F}^{a, \bar a}_{\zeta}(X)$  
 appear only through the difference ${\cal F}^{a}_{\zeta}(X)-
{\cal F}^{ \bar a}_{\zeta}(\zeta - X)$. 
In  a recent paper \cite{mgb}, Golec-Biernat and Martin 
argued that the decomposition of $\tilde {\cal F}^a_{\zeta}  (X)$ 
in the middle region 
 into  ${\cal
F}^{a}_{\zeta}(X)$ and ${\cal F}^{ \bar a}_{\zeta}(\zeta - X)$ 
parts made  in Ref.\cite{npd}  
amounts to ``doubling the quark degrees of freedom''
\footnote{They also proposed to split   our function  
$ \tilde {\cal F}^{a}_{\zeta}(X)$ into overlapping 
$0 \leq X \leq 1$ and 
$-1+ \zeta \leq X \leq \zeta$ parts to 
introduce ``off-diagonal'' 
``quark'' $ \hat {\cal F}^{ a}_{\zeta}  (X) =  
\theta(0 \leq X \leq 1) \tilde {\cal F}^{a}_{\zeta}(X)$
and ``antiquark''  $ \hat {\cal F}^{\bar a}_{\zeta}  (\zeta - X)
= - \theta(0 \leq X \leq 1) \tilde {\cal F}^{a}_{\zeta}(\zeta - X)$
distributions both of which include  the same middle part
of  $ \tilde {\cal F}^{a}_{\zeta}(X)$. }.  
 Compared to Ref. \cite{npd}, we have  
an extra  function
  $ {\cal F}^{M}_{\zeta}  (X)$ in Eq.(\ref{three}), so  one may 
  question now whether it make sense to 
represent  $\tilde {\cal F}^a_{\zeta}  (X)$ 
as a sum of  three functions in the $0 \leq X \leq \zeta$ region.
Of course, if there were  only one value of 
$\zeta$ in the nature,   one would never get an  idea about 
how much of $\tilde {\cal F}^a_{\zeta}  (X)$
should be attributed  to ${\cal F}_{\zeta}^{a}(X), 
{\cal F}_{\zeta}^{\bar a}(X) $ 
or ${\cal F}_{\zeta}^{M}(X)$.
The crucial missing element is the interplay
between $\zeta$ and $X$ dependences.
We stress  that 
our decomposition of 
$\tilde {\cal F}^a_{\zeta}  (X)$
is based on the  splitting of 
the underlying  $y$-DDs $F^a(x,y)$
into positive-$x$, negative-$x$ and zero-$x$ parts.
The   DDs contain information about NFPDs 
for all possible $\zeta$'s and $X$'s, and   that is why 
the DDs produce an unambiguous
 decomposition:  DDs ``know'' not only 
what is the shape of  $\tilde {\cal F}^a_{\zeta}  (X)$
 for a particular  $\zeta$, but also 
how this shape  would change   if one would 
 take another $\zeta$. The simplest illustration
 of  interplay between 
 $X$ and $\zeta$ dependences 
 is provided by NFPDs ${\cal F}_{\zeta}^{M}(X)=
 \theta (0\leq X/\zeta  \leq 1)
 \varphi (X/ \zeta) / |\zeta|$
 corresponding to singular parts of DDs.
 Clearly,  knowing  ${\cal F}_{\zeta}^{M}(X)$ 
  at some $\zeta = \zeta_0$,
 we can obtain its shape  for any other 
 $\zeta$ by rescaling. 
 To demonstrate that NFPDs with such a behavior
 can be obtained only from singular DDs, 
 we  write
 a formal inversion 
 of the basic relation  (\ref{71}) 
\begin{equation}
F(x,y) = \int_{-\infty}^{\infty} dX \int_{-\infty}^{\infty}
\Delta [X- x - \zeta y] \, {\cal F}_{\zeta} (X) \, d \zeta \ , 
\label{inverse} 
\end{equation}
where   the (mathematical) distribution 
$\Delta (z)$ is defined by
\begin{equation}
\Delta [z] = \frac1{(2 \pi)^2} \int_{-\infty}^{\infty}
 |\mu | \, e^{i \mu z } d \mu  \, . \label{Delta}
 \end{equation}
Taking  $ {\cal F}_{\zeta}^M (X) 
=\theta(0\leq X/\zeta  \leq 1) \varphi (X/\zeta)/|\zeta| $
and using the following property  of the $\Delta$-function
 \begin{equation}
 \int_{-\infty}^{\infty} \Delta [a - \zeta b] \,
 d \zeta = \delta (a) \delta (b)  \, , \label{ddelta}
 \end{equation} 
  we obtain from Eq. (\ref{inverse})  that 
$F_M (x,y) = \delta (x) \varphi (y)$.

Thus, information contained in SPDs originates from two  
physically different sources: meson-exchange type contributions  
${\cal F}_{\zeta}^M(X)$ 
coming from the singular $x=0$ parts of DDs
and  the functions 
${\cal F}_{\zeta}^a(X)$,  ${\cal F}_{\zeta}^{\bar a}(X)$
 obtained by scanning the $x \neq 0$  parts of  
DDs $F^a(x,y)$, $F^{\bar a} (x,y)$. 
The support of exchange contributions is restricted 
to $0 \leq X \leq \zeta$. Up to rescaling, the function
${\cal F}_{\zeta}^M(X)$  has the same shape for all $\zeta$.
For any nonvanishing $X$, these exchange terms  become  invisible 
in the forward limit $\zeta \to 0$.  
On the other hand, 
interplay between $X$ and $\zeta$ 
dependences  of the functions 
${\cal F}_{\zeta}^a(X)$,  ${\cal F}_{\zeta}^{\bar a}(X)$
is quite nontrivial and their 
support in general covers the whole   $0\leq X \leq 1$ region
for all $\zeta$ including the forward limit 
$\zeta =0$ in which they convert into 
 the usual (forward) densities
$f^a(x)$, $f^{\bar a}(x)$. The latter 
are   
rather well known  from inclusive measurements. Hence,  
 information  contained in $f^a(x)$, $f^{\bar a}(x)$
can be used to 
restrict the models for 
${\cal F}_{\zeta}^a(X)$,  ${\cal F}_{\zeta}^{\bar a}(X)$.
Note that the functions $F^a(x,y)$ and $F^{\bar a} (x,y)$ 
 are  independent  as are their $\zeta$-sensitive  
scans ${\cal F}_{\zeta}^a(X)$ and  ${\cal F}_{\zeta}^{\bar a}(X)$.
Instead of  $F^a(x,y)$ and $F^{\bar a} (x,y)$, one can use as 
independent functions their sum 
$F^a(x,y)+F^{\bar a} (x,y)$ which contributes to the quark singlet
functions and the difference  $F^a(x,y) - F^{\bar a} (x,y)$ which 
appears in the valence functions.  
Extending the  DDs onto the whole $-1 \leq x \leq 1$ 
segment does not require extra dynamical 
information:  one should only take into account 
that the singlet term $\tilde f^S_a (x,\alpha)$
must  be  odd in $x$ (see Eq.(\ref{singlet}))  while the valence 
term $\tilde f^V_a (x,\alpha)$ must  be  
even in $x$ (see Eq.(\ref{valence})). 
As a result, the singlet contribution $\tilde {\cal F}_{\zeta}^{S,a}(X)$ 
is an odd function of $X-\zeta/2$ while the valence one 
$\tilde {\cal F}_{\zeta}^{V,a}(X)$  is an even function 
of  $X-\zeta/2$.

In our approach, DDs are the starting point
while SPDs are derived from them by integration.
However, even if one starts directly 
with SPDs,  the latter possess a property
which forces  
the use of double distributions. 
According to Eq.(\ref{71}), the  
$X^N$ moment of  ${\cal F}_{\zeta} (X)$
{\it must be  a polynomial} 
 in $\zeta$ of a degree not larger than $N$.
A similar  statement holds 
for  off-forward distributions 
$\tilde H(\tilde x,\xi;t)$:  their $\tilde x^N$
moments are $N$th order polynomials of $\xi$.  
As explained  by X. Ji  \cite{jirev},  
this restriction on the interplay between
$\tilde x$ and $\xi$ dependences of $\tilde H(\tilde x,\xi;t)$ 
follows from  a simple fact that  the Lorentz  indices
$\mu_1 \ldots \mu_N$ of the 
 nonforward 
matrix elements of a local operator $O^{ \mu_1 \ldots \mu_N}$
can be carried either by $P^{\mu_i}$ or by $r^{\mu_i}$.
As a result, 
\begin{equation}
\langle P-r/2 | \phi (0) (\stackrel{
\leftrightarrow}{
\partial^+} )^N
\phi(0)|P+r/2\rangle =
\sum_{k=0}^{N}{N \choose k} (P^+)^{N-k} (r^+)^{k}  A_{Nk} \, = 
(P^+)^{N}\sum_{k=0}^{N} {N \choose k}\xi^k    A_{Nk} \, ,
\label{poly} 
\end{equation}
where ${N \choose k} \equiv N!/(N-k)!k!$ 
is the combinatorial coefficient.
Our  derivation (\ref{offfor}) of 
OFPDs from  $\alpha$-DDs
automatically satisfies the 
polynomiality condition (\ref{poly}), since 
\begin{equation}
\int_{-1}^1 \tilde H(\tilde x,\xi;t)\, \tilde x^N \, d \tilde x=
\sum_{k=0}^{N} \, \xi^k \, {N \choose k} \, 
\int_{-1}^1 dx \int_{-1 +|x|}
^{1-|x|} \tilde f(x,\alpha) \,  x^{N-k} \alpha^k
\, d\alpha \  .  \end{equation}
Hence, the coefficients $A_{Nk}$ in (\ref{poly}) 
are given by  double moments
of $\alpha$-DDs. This  means that modeling 
SPDs one cannot choose the coefficients $A_{Nk}$
arbitrarily: symmetry and support properties of DDs dictate  
a nontrivial interplay between 
$N$ and $k$ dependences of $A_{Nk}$'s.  
 After this observation,   
the use of DDs is an unavoidable step
in  building consistent parametrizations of SPDs.

The formalism of  DDs  also allows one to easily establish 
some important  properies of skewed distributions.
Notice  that due to the cusp at the  upper corner
of the DD-life triangle, 
the length of the integration 
line nonanalytically depends on $X$ for $X=\zeta$. 
Hence, unless a double distribution 
identically  vanishes  in a finite region 
around  the upper corner  of the DD support triangle,
the $X$-dependence of the relevant nonforward distribution 
{\it must be nonanalytic}  at the border point $X=\zeta$. 
Furthermore,  the length of the 
integration line vanishes when  $X \to 0$.
As a result, the components ${\cal F}_{\zeta}^{a,\bar a} (X)$ 
vanish at $X=0$
if the relevant double distribution $F^{a,\bar a} (x,y)$ is not too
singular for small $x$. The combined contribution 
of ${\cal F}_{\zeta}^a (X)$ and ${\cal F}_{\zeta}^{\bar a}
 (\zeta -X)$ into the total function 
$\tilde {\cal F}_{\zeta}^a (X)$ 
in this case is continuous at the nonanalyticity  
points $X=0$ and $X=\zeta$.
As emphasized in Ref. \cite{npd},  because 
of the  $1/X$ and $ 1/(X-\zeta)$ factors 
($1/(\tilde x \pm \xi)$ factors if OFPD formalism is used) 
contained in  hard amplitudes, this property  is 
crucial for  pQCD factorization 
in  DVCS and other
hard electroproduction processes.  
Note,  that there is also the exchange contribution
${\cal F}_{\zeta}^M (X)$. 
If  it comes from a $\delta (x) \varphi (y)$ type term and 
$\varphi (y)$ 
 vanishes at the end-points $y=0,1$  the   
${\cal F}_{\zeta}^M (X)$ part of
NFPD  vanishes at $X=0$ and $X=\zeta$.
The total function 
$\tilde {\cal F}_{\zeta}^a (X)$ 
is  then continuous at these nonanalyticity points
(OFPDs $\tilde H (\tilde x, \xi;t) $ in this case 
are continuous at $x = \pm \, \xi$).
In the quark singlet case, the DDs should be odd in $x$,
hence the singular term involves $\delta^{\prime} (x) \varphi(y)$
(or even higher odd derivatives of $\delta (x)$). 
One can  get a continuous SPD in this case only 
if  $\varphi ^ {\prime} (y)$  vanishes at the end points.
Such  a  restriction  might be  too   strong to be satisfied in all 
cases. 
In particular,  
an   essentially discontinuous behavior
of  singlet quark OFPDs for $\tilde x = \pm \, \xi$ 
was obtained in  a  nonperturbative
(chiral soliton)   model \cite{ppp}.

\section{Models for skewed distributions}

The properties discussed above can be illustrated by
SPDs  constructed using    simple 
models of DDs specified in  Section III.
In particular, for the model \mbox{$F^{(0)}(x,y) = \delta (y -\bar x/2)
f(x)$} (equivalent to  \mbox{$f^{(0)}(x,\alpha) = \delta (\alpha)
f(x)$} ), 
we get 
\begin{equation}
{\cal F}_{\zeta}^{(0)} (X) = \frac{\theta(X \geq \zeta/2)}{1-\zeta/2}
 f \left (\frac{X-\zeta/2}{1-\zeta/2} \right ) \, ,
 \label{model}
  \end{equation}
i.e.,  NFPDs for non-zero  $\zeta$ are obtained from 
the forward distribution $f(X)\equiv {\cal F}_{\zeta=0} (X)$  
 by   shift and rescaling. This is an example of  a peculiar 
case of a  DD with  an empty upper corner:  it gives NFPDs with 
no explicit nonanalyticity at $X=\zeta$. ``As a compensation'',
$ {\cal F}_{\zeta}^{(0)}(X) $ vanishes not only for $X=0$,
but on the finite segment $0 \leq X \leq \zeta/2$.
  Using the relations  \cite{npd} 
\begin{equation} 
\tilde H (\tilde x,\xi; t)|_{\tilde x > \xi}
=  {(1-\zeta/2)} \, {\cal F}_{\zeta} (X;t)|_{X>\zeta}  \  \  ;
 \  \     \tilde x =  \frac{X-\zeta/2}{1-\zeta/2}  \  \  ;  \  \  
\xi = \frac{\zeta}{2- \zeta}   \label{onfpd}
 \end{equation}
between our nonforward distributions in the $X> \zeta$ region
and 
  Ji's off-forward parton distributions  $ H(\tilde x,\xi;t)$
\cite{ji}  in the $\tilde x > \xi$ region,  
 one can see that the narrow $F^{(0)}(x,y)$ 
ansatz gives the simplest    model 
  $H^{(0)}(\tilde x,\xi; t=0) = f(x)$ in which  OFPDs at $t=0$
   have no $\xi$-dependence.
 This result can be obtained directly 
 by using the model $f^{(0)}(x,\alpha) = \delta (\alpha)
f(x)$ for the $\alpha$-DDs.
Another example is the model \cite{ffgs,maryskin}
in which  NFPDs do not
depend on $\zeta$, i.e., $ {\cal F}_{\zeta} (X) =f(X)$.
Using the inversion formula
(\ref{inverse}) and Eq. (\ref{ddelta}),
we obtain $F(x,y) = \delta (y) f(x)$, $i.e.,$ 
the support of this  DD is on the $y$-axis only,
which violates  the mandatory
$y \leftrightarrow 1-x-y$ symmetry.
Unlike the $\xi$-independent ansatz 
for OFPDs, the $\zeta$-independent ansatz for NFPDs 
is forbidden.

In case of  two other models, simple  analytic results 
can be obtained only for some   explicit forms of  $f(x)$.
For the ``valence quark''-oriented ansatz $\tilde f^{(1)}(x,\alpha)$,
the following choice 
\begin{equation}  f^{(1)}(x) = 
 A \, 
  x^{-a} (1-x)^3  \label{74} \end{equation}
is   both  close 
to phenomenological   quark distributions
and   produces a simple expression
for the double distribution since the denominator
$(1-x)^3$ factor in Eq. (\ref{mod123}) is canceled.
As a result, the integral in Eq. (\ref{offfor})
is easily performed and   we get
\begin{equation}
\tilde H^{(1V )}(\tilde x, \xi)|_{|\tilde x| \geq \xi}  =
\tilde A \left \{ \left[ (2-a) \xi (1- \tilde x) (x_1^{2-a} + x_2^{2-a}) +
(\xi^2 -\tilde x)(x_1^{2-a} - x_2^{2-a}) \right ] \, \theta (\tilde x) 
+ ( \tilde x \to -\tilde x) \right \}  \label{outs} 
\end{equation}
for  $|\tilde x |\geq \xi$,  where $\tilde A =
3A \,  \Gamma(1-a)/2\Gamma(4-a)$, and 
\begin{equation}
\tilde H^{(1V) }(\tilde x, \xi)|_{|\tilde x| \leq \xi}  =
\tilde A \left \{ x_1^{2-a}[(2-a) \xi (1- \tilde x) +
(\xi^2 -x)] + ( \tilde x \to -\tilde x) \right \} \label{middles}
\end{equation}
in the middle $ -\xi \leq \tilde x \leq \xi$ region.
We use here  the notation $x_1=(\tilde x + \xi)/(1+\xi)$
  and 
$x_2=(\tilde x - \xi)/(1-\xi)$ \cite{jirev}.
As expected, these expressions are explicitly non-analytic 
for $x = \pm \xi$. 
It is interesting to note that 
in a particular case  $a=0$, the 
$x>\xi$ part of OFPD has the same $x$-dependence 
as its forward limit, differing from it by an overall $\xi$-dependent 
factor only:
\begin{equation}
\tilde H^{(1V) }(\tilde x, \xi)|_{a=0} = 
 A \, \frac{(1-|\tilde x|)^3}{(1-\xi^2)^2} 
\, \theta (|\tilde x| \geq \xi) \, 
+ A \, \frac{\xi +2 -3 \tilde x^2/\xi}{2(1+\xi)^2} \, 
 \theta (|\tilde x| \leq \xi)
\, .    \label{(1-x)^3}
\end{equation} 
To extend this expression onto negative values of 
$\xi$, one should  substitute $\xi$ by $|\xi|$.
One can check, however, that no odd powers of $|\xi|$ 
would appear in the $\tilde x^N$ 
moments of $\tilde H^{(1V) }(\tilde x, \xi)$.

For  the singlet quark distribution, the  $\alpha$-DDs
$\tilde f^S( x, \alpha)$ should be odd functions 
of $x$. Still, we can use  the model like (\ref{74}) for 
the $x>0$ part, but 
take $\tilde f^{(1S)} ( x, \alpha)|_{x \neq 0} 
=  f^{(1)}( |x|, \alpha)\, {\rm sign} (x)$. 
Note, that the integral (\ref{offfor})  producing $\tilde H^S(\tilde x, \xi)$ 
in the $|\tilde x| \leq \xi$ region
would diverge  for  $\alpha \to \tilde x /\xi$  
 if $a \geq  1$, which is the usual case 
 with standard parametrizations of singlet quark 
 distributions at  sufficiently large $Q^2$. 
However, due to the antisymmetry of  $\tilde f^S( x, \alpha)$
wrt  $x \to -x$ and its symmetry wrt $\alpha \to -\alpha$,
the singularity at 
$\alpha = \tilde x /\xi$ can be  integrated  
using the principal value 
prescription   which in this case 
produces the $x\to -x$ antisymmetric version 
of Eqs.(\ref{outs}) and (\ref{middles}). 
As far as $a<2$, the resulting functions
 are finite for all
$\tilde x$ and  continuous at $\tilde x = \pm \xi$.
 For $a=0$, 
the  middle part  reduces to 
\begin{equation}
\tilde H^{(1S)} (\tilde x, \xi)|_{|\tilde x| \leq \xi, a=0} = 
A \,  x\, \frac{3 \xi^2 -2 x^2 |\xi| - x^2}{2|\xi|^3 (1+|\xi|)^2} 
  \,  .
\end{equation}
 
Evidently, the use of the principal value prescription
 is equivalent to imposing a subtraction 
 procedure for  the divergent   
second integral in Eq. (\ref{710}) defining  the 
untilded functions $H^{a, \bar a}(\tilde x, \xi)$.

 In general case, to study  the deviation of skewed distributions  
 from their forward counterparts for small $\xi$ (or $\zeta$), 
let us consider 
 the integral producing the 
$x \geq \xi$ part of $H(x,\xi)$
[see   Eq.(\ref{710})]  and expand it in powers of $\xi$:
\begin{equation}
H(\tilde x;\xi)|_{\tilde x \geq \xi} 
 = f(\tilde x) +  \xi^2 \left [ 
\frac12 \int_{- (1- \tilde x)}^{(1- \tilde x)}
\frac{ \partial^{2}f (\tilde x,  \alpha  )}{\partial \tilde x^{2}}
 \,\alpha  ^{2} \, d  \alpha   \,  + (1- \tilde x)^2
\left. \left (\frac{ \partial f (\tilde x,  \alpha  )}
 {\partial \alpha } 
 -2 \frac{ \partial f (\tilde x,  \alpha  )}
 {\partial \tilde x  } \right ) \right |_{\alpha = 1 - \tilde x}
 \right ] +  \ldots \, , 
\label{exxi2} 
\end{equation} 
where  $f(\tilde x)$ is  the forward distribution.
For small $\xi$, the corrections 
are formally $O(\xi^2)$, i.e., they look very small.
However,   if $f(x,\alpha)$  
has a singular 
behavior like $x^{-a}$, then 
$$\frac{ \partial^{2} f(\tilde x,  \alpha  )}{\partial \tilde x^{2}}
\sim \frac{a (1+a)}{\tilde x^2} f(\tilde x,  \alpha  )\ , $$
and the relative suppression of the first correction 
is $O(\xi^2/\tilde x^2)$.   The corrections 
are tiny for all $\tilde x$ except for the region  $\tilde x \sim \xi$
 where the correction has   no parametric smallness.
 Nevertheless, even in this region it is suppressed 
numerically, because the $\alpha  ^2$ moment is rather small
for   a distribution concentrated in the small-$\alpha  $ region.
 It is easy to write expicitly all  the   terms 
 which are not suppressed in the $\tilde x \sim \xi \to 0$ limit
  \begin{equation}
H(\tilde x;\xi) = \sum_{k=0} 
\frac{\xi^{2k}}{(2k)!}   
 \int_{- 1}^{1}
\frac{ \partial^{2k} f (\tilde x,  \alpha  )}{\partial \tilde x^{2k}}
 \,\alpha  ^{2k} \, d  \alpha   \, + \ldots
  \, ,
 \end{equation}
 where the ellipses denote the terms vanishing in this  
 limit.
 Due to strong  numerical suppression of higher terms,
 the series converges rather fast. 
 For small $x$, we can neglect the $x$-dependence of the profile
 function $h(x,\alpha)$ in Eq. (\ref{65n})
 and  take the model $f(x,\alpha) = f(x) \rho (\alpha)$
 with $\rho (\alpha)$ being a symmetric weight 
 function on $-1 \leq \alpha \leq 1$ whose integral over $\alpha$ 
 equals 1. In the region where both  $\tilde x$ and $\xi$ are small,
 we can approximate Eq. (\ref{offfor}) by 
 \begin{equation}
\tilde H(\tilde x;\xi) =    
 \int_{- 1}^{1} 
\tilde  f (\tilde x -\xi   \alpha  ) \rho (\alpha) \, d \alpha
  + \ldots \, ,
 \end{equation}
 i.e., the OFPD $H(\tilde x;\xi)$ is obtained 
 in this case by averaging  the  usual (forward) 
 parton density $\tilde f(x)$ 
(extended onto $-1 \leq x \leq 1$) over the region 
 $\tilde x - \xi \leq x \leq \tilde x+\xi$ 
 with the weight $\rho (\alpha)$. 
 In terms of NFPDs, the relation is 
  \begin{equation}
 \tilde {\cal F}_{\zeta} (X )  = 
 \int_{- 1}^{1} 
\tilde  f (X- \zeta (1+   \alpha)/2  ) \rho (\alpha)
 \, d \alpha
 + \ldots \, ,
 \end{equation}
 i.e., the average is taken over the region 
 $X -\zeta \leq x \leq X$.
 
 The imaginary part of  hard exclusive 
 meson electroproduction amplitude is determined by
 the skewed distributions at 
 the border point. For this reason, the magnitude 
 of  ${\cal F}_{\zeta} (\zeta)$ [or $H(\xi, \xi)$],
   and its relation to the  forward densities
 $f(x)$   has a practical interest.
 Assuming   the infinitely narrow weight 
 $\rho(\alpha) = \delta (\alpha)$,
 we have ${\cal F}_{\zeta} (X ) = f(X-\zeta/2) + \ldots $ 
 and $H(x,\xi) = f(x)$.
 Hence, both ${\cal F}_{\zeta} (\zeta)$ and  $H(\xi, \xi)$
 are given by $f(x_{Bj}/2)$ because  $\zeta = x_{Bj}$
 and $\xi =x_{Bj}/2 +\ldots$.  Since the argument
 of $f(x)$ is twice smaller than in deep inelastic scattering,
this results in an enhancement factor. In particular, if 
  $f(x) \sim x^{-a}$ for small  $x$, the ratio
  ${\cal F}_{\zeta} (\zeta ) /f(\zeta )$ is 
  $2^a$. 
 The use of a wider weight function $\rho (\alpha)$ produces further
 enhancement. For example, taking 
 $\rho (\alpha)= \frac34 (1-\alpha^2)$ and
$f(x) \sim x^{-a}$ we get 
 \begin{equation}
 \frac{{\cal F}_{\zeta} (\zeta ) }{f(\zeta )}
  = \frac1{(1-a/2)(1-a/3)}
 \end{equation}
 which is larger than $2^a$ for $0< a <2$. 
Due to evolution, the effective parameter $a$ 
is an increasing function of $Q^2$.
As a result, the above ratio slowly increases with $Q^2$.

Finally, I want to point out
that possible profiles of $f(x,\alpha)$ 
in the $\alpha$-direction
are restricted by inequalities 
(see \cite{maryskin,jirev,ddee,pisoter}) relating
skewed and forward distributions. For quark OFPDs, I obtained 
 \cite{ddee} 
 \begin{equation}
 H^q(\tilde x,\xi)  \leq \sqrt{ \frac1{1-\xi^2} \,
 f \left ( \frac{\tilde x+\xi}{1+\xi} \right )
 f \left ( \frac{\tilde x-\xi}{1-\xi} \right )} \, 
= \frac1{\sqrt{ 1-\xi^2}} \sqrt{f(x_1) f(x_2)} \, . 
\label{Hq}
 \end{equation}
 If one uses   the infinitely narrow
model $f^{(0)}(x,\alpha) = f(x)\,  \delta (\alpha)$ 
[corresponding to $H^{(0)}(\tilde x,\xi) = f(\tilde x)$],
 the inequality (\ref{Hq})  is satisfied  
for any function $f(x)$ of
$x^{-a}(1-x)^b$ type  with $a \geq 0, \, b>0$.
For the  model (\ref{(1-x)^3}) which has a 
wider $\frac34 (\bar x^2 - \alpha^2)$
profile and $f(x)=A \, (1-x)^3$, 
the inequality (\ref{Hq}) is exactly saturated.
If one  takes the model 
$f^{(4)}(x,\alpha) = \bar x  \, f(x) \,  \delta(\bar x^2 - \alpha^2)$
with an extremely wide
profile, one obtains the result 
$H^{(4)}(\tilde x,\xi) = 
\frac12 \{ f(x_1)/(1+\xi) + f(x_2)/(1-\xi)\} $
which violates (\ref{Hq}).

\section{Summary}

In this paper, we treated 
double distributions  as 
the basic  objects for  parametrizing  
nonforward matrix elements.  
An alternative description
in terms of skewed distributions was 
 obtained 
by  an appropriate integration
of  relevant DDs. The use 
of DDs helps to 
establish   important features
of SPDs such as their   nonanalyticity  at 
the border points $X=\zeta$ and $\tilde x = \pm \xi$.
DDs are  crucial for securing the 
property that 
the  moments of SPDs should be   polynomial
in the skewedness parameter. 
For these reasons, the use 
of DDs is unavoidable in constructing   
 consistent models 
of SPDs.

\section{Acknowledgements}

I acknowledge stimulating discussions  and communication 
with I.I. Balitsky,  A.V. Belitsky, S.J. Brodsky,
J.C. Collins,  L.L. Frankfurt,
 K. Golec-Biernat, X. Ji, L. Mankiewicz, A.D. Martin, 
 I.V. Musatov, G. Piller, M.V.  Polyakov,
  M.G.  Ryskin, A. Sch\"afer, A. Shuvaev, 
  M.A.  Strikman, O.V. Teryaev and 
  C. Weiss. 
 This work was supported by the US 
 Department of Energy under contract
DE-AC05-84ER40150.


\begin{thebibliography}{99}

\bibitem{ji}  X. Ji, { Phys.Rev.Lett.} {\bf 78}  610
(1997);{ Phys.Rev. } D {\bf 55} (1997) 7114.
\bibitem{compton}  A.V. Radyushkin,  { Phys. Lett.} B {\bf 380}
(1996) 417;   {Phys.  Lett.} B {\bf 385} (1996) 333.
\bibitem{npd} A.V. Radyushkin, {Phys. Rev. } D {\bf 56} (1997) 5524.
\bibitem{cfs} J.C. Collins,
L. Frankfurt and M. Strikman,  {Phys. Rev. } D {\bf 56} (1997) 2982.
\bibitem{drm} D. M\"uller, D. Robaschik, B. Geyer, 
F.-M. Dittes and J. Ho{\v r}ej{\v s}i,
{ Fortschr.Phys.} {\bf 42} (1994) 101.
 \bibitem{spectral}  A.V. Radyushkin,  {Phys. Lett.}   
B {\bf 131} (1983) 179. 
\bibitem{christian} C. Weiss, talk at the ECT$^*$ Workshop
``Coherent QCD processes with nucleons and nuclei'',
Trento, Italy,  September 1998.
\bibitem{lech} L. Mankiewicz, G. Piller and  T. Weigl, 
  Eur.  Phys. J. C {\bf 5}  (1998) 119. 
  \bibitem{realco} A.V. Radyushkin, hep-ph/9803316,   Phys. Rev. D
{\bf 58} (1998) 114008.
\bibitem{jirev} X.Ji,  J. Phys. G {\bf 24} (1998) 1181.
\bibitem{mgb}  K. Golec-Biernat and A.D. Martin,
hep-ph/9807497.
\bibitem{ppp} V.Yu. Petrov, P.V. Pobylitsa, 
M.V. Polyakov, I. Bornig, K. Goeke
and  C. Weiss,  {Phys.  Rev.} D {\bf 57}
(1998) 4325.
\bibitem{ffgs} L. Frankfurt, A. Freund, V. Guzey and 
 M. Strikman, {Phys.  Lett.} B {\bf 418} (1998) 345.
  \bibitem{maryskin}  A.D. Martin and  M.G. Ryskin, 
 Phys. Rev. D {\bf 57} (1998) 6692.
\bibitem{ddee}  A.V. Radyushkin, hep-ph/9805342,
 to appear in Phys. Rev. D.
\bibitem{pisoter} B. Pire,  J. Soffer and O.V. Teryaev, hep-ph/9804284.





\end{thebibliography}
\end{document}